\documentclass[fleqn,twoside]{article}
\usepackage[headings]{espcrc2}

\readRCS
$Id: espcrc2.tex,v 1.2 2004/02/24 11:22:11 spepping Exp $
\usepackage{graphicx}

\newcommand{\AmS}{{\protect\the\textfont2
  A\kern-.1667em\lower.5ex\hbox{M}\kern-.125emS}}

\title{Sum rules of polarized photon structure functions revisited
\thanks{Presented by K. Sasaki} \\
--- NNLO corrections to the first moment of $g_1^\gamma (x,Q^2,P^2)$ ---}

\author{T. Ueda\address[YNU]{Department of Physics, Faculty of Engineering, 
             Yokohama National University, \\ 
        \ Yokohama 240-8501, Japan},
        T. Uematsu\address{Department of Physics, Faculty of Science, Kyoto University, \\
        \ Kyoto 606-8501, Japan}
        and
        K. Sasaki\addressmark[YNU]}
       
\runtitle{NNLO corrections to the first moment of $g_1^\gamma (x,Q^2,P^2)$}

\begin{document}

\begin{abstract}
We present the next-to-next-to-leading order ($\alpha \alpha_s^2$) corrections to the 
first moment of the  polarized virtual photon 
structure function $g_1^\gamma(x,Q^2,P^2)$ in the kinematical region
$Q^2\gg P^2 \gg \Lambda^2$ in QCD. We find that the 
$\alpha \alpha_s^2$ corrections are about 3\%
of the sum of the leading ($\alpha$) and the next-to-leading ($\alpha\alpha_s$) contributions,
when $Q^2=30\sim 100 {\rm GeV}^2$ and $P^2=3 {\rm GeV}^2$.
\vspace{1pc}
\end{abstract}

\maketitle

\section{Introduction}

\begin{picture}(0,0)%
  \put(350,420){%
    \put(2.3,-75){YNU-HEPTh-06-103}
    \put(2.3,-90){KUNS-2017}
    \put(2.3,-105){April 2006}
  }%
\end{picture}%
The investigation of the photon structure has been an active field
of research both theoretically and experimentally~\cite{Kraw,Nisi,Klas,Schi}.
Also there has been growing interest in the study of the spin 
structure of photon. In particular, the first moment of the polarized 
photon structure function $g_1^\gamma$ has attracted attention 
in connection with its relevance to the QED and QCD axial
anomaly~\cite{ET,BASS,NSV,FS,BBS}. The polarized photon structure functions 
can be measured from two-photon
processes in the polarized $e^+e^-$ collider experiments as shown in Fig.~1, where $-Q^2
(-P^2)$ is the  mass squared of the probe (target) photon.

\begin{figure}
\begin{center}
\includegraphics[scale=0.7]{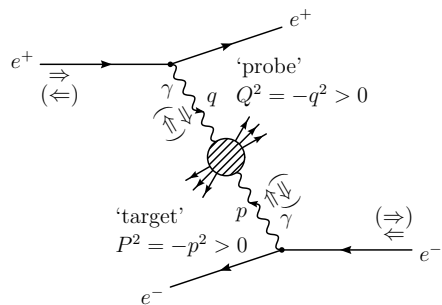}
\vspace{-0.5cm}
\caption{Deep inelastic scattering on a virtual photon in $e^+~e^-$ collision.}
\end{center}
\end{figure}

For a real photon ($P^2=0$) target, there exists only one 
spin-dependent structure function $g_1^\gamma(x, Q^2)$. The QCD analysis 
for $g_1^\gamma$ was performed in the leading order (LO) \cite{KS} and 
in the next-to-leading order (NLO) \cite{SV,GRS}. In the case of a virtual 
photon target ($P^2\neq 0$) there appear two spin-dependent structure functions,  
$g_1^\gamma(x, Q^2, P^2)$ and $g_2^\gamma(x, Q^2, P^2)$. The former has been 
investigated up to the NLO in QCD by the present authors in\cite{SU1,SU2}, and also
in the second paper of \cite{GRS}. In Ref.\cite{SU1} the
structure function $g_1^\gamma(x, Q^2, P^2)$ was analysed in the kinematical region
\begin{equation}
\Lambda^2\ll P^2 \ll Q^2~, \label{Kinematical}
\end{equation}
where $\Lambda$ is the QCD scale  parameter. The
advantage to study the virtual photon target in this kinematical region is that 
we can calculate structure functions  by the
perturbative method without any experimental data input~\cite{UW}, which is contrasted
with the case of the real photon  target where in the NLO there exist
nonperturbative pieces. 

In this talk we focus on the photon structure function $g_1^\gamma$ and, in particular, 
on the sum rule and present the
result  of the NNLO ($(\alpha\alpha_s^2)$) corrections to the 
first moment of $g_1^\gamma(x,Q^2,P^2)$ in the case of the kinematical region 
(\ref{Kinematical}).

\section{The sum rule of $g_1^\gamma$}
The polarized structure function $g_1^\gamma$ of the real photon
satisfies a remarkable sum rule \cite{ET,BASS,NSV,FS,BBS}
\begin{equation}
\int_0^1 dx~ g_1^\gamma(x,Q^2)=0~.\label{g1gammaReal}
\end{equation}
In fact, the authors of Ref.~\cite{BBS} 
showed that the sum rule (\ref{g1gammaReal}), holds to all orders  
in perturbation theory in both QED and QCD.

When the target photon becomes off-shell, i.e., $P^2\not= 0$, the first
moment of the corresponding photon structure function
$g_1^\gamma(x,Q^2,P^2)$ does not vanish any more.  Indeed, for the case 
$\Lambda^2\ll P^2 \ll Q^2$, the first moment has been calculated 
up to the NLO ($O(\alpha\alpha_s)$) in QCD as follows~\cite{NSV,SU1};
\begin{eqnarray}
&&\hspace{-0.8cm}\int_0^1dx g_1^\gamma(x,Q^2,P^2)\nonumber \\
&=&\! \!-\frac{3\alpha}{\pi}
\left[\sum_{i=1}^{n_f}e_i^4\left(1-\frac{\alpha_s(Q^2)}{\pi}\right)\right.
\nonumber\\
&&\! \!-\left.\frac{2}{\beta_0}(\sum_{i=1}^{n_f}e_i^2)^2\left(
\frac{\alpha_s(P^2)}{\pi}-\frac{\alpha_s(Q^2)}{\pi}\right)\right],\label{Oalpha}
\end{eqnarray}
with $\beta_0\!=\!11\!-\!2n_f/3$ being the one-loop QCD $\beta$ function. 
Here 
$\alpha\!=\!e^2/4\pi$ and $\alpha_s(Q^2)\!=\!{\overline g}^2(Q^2)/4\pi$
are the QED coupling constant and the QCD running coupling constant,
respectively, and
$e_i$ is the electromagnetic charge of the active quark with flavour $i$ in the unit of 
proton charge and $n_f$ is the number of active quarks.  
Note that the r.h.s. of (\ref{Oalpha}) does not involve any
experimental data input. This is because  the nonperturbative pieces, which
take part in the real photon target case, can be neglected 
in the kinematical region (\ref{Kinematical}) and the whole twist-two 
contributions to $g_1^\gamma(x,Q^2,P^2)$ can be computed by the perturbative 
method. It is also noted that the first term in the square brackets of the r.h.s. of 
(\ref{Oalpha})  results from the QED triangle anomaly while the second term comes from 
the QCD triangle anomaly.

\section{The NNLO ($\alpha\alpha_s^2$) corrections}

Now we calculate the NNLO $(\alpha\alpha_s^2)$ corrections to the 
first moment of $g_1^\gamma(x,Q^2,P^2)$.
First recall that for the operator product expansion (OPE) of two electromagnetic 
(and thus gauge-invariant) currents,   only gauge-invariant operators need be
included with their renormalization basis~\cite{Collins}.
Since there is no gauge-invariant twist-two  gluon and photon 
operators with  spin one, we need to consider only quark operators, i.e.,
the flavour singlet $R_{S}^{\sigma}$ 
and nonsinglet $R_{NS}^\sigma$ axial currents, as follows: 
\begin{equation}
  R_{S}^{\sigma} =\overline{\psi}\gamma^\sigma\gamma_5~1\psi       
, \ \ 
  R_{NS}^\sigma =\overline{\psi}\gamma^\sigma\gamma_5
             (Q^2_{ch}-\langle e^2 \rangle 1)\psi, \label{axialcurrent}
\end{equation}
where $1$ is an $n_f \times n_f$ unit matrix and $Q^2_{ch}$ is the square of the
$n_f \times n_f$ quark-charge matrix.


Then the first moment of $g_1^\gamma(x,Q^2,P^2)$ is expressed as   
\begin{eqnarray}
&&\hspace{-0.8cm}\int_0^1 dx g_1^\gamma(x,Q^2,P^2)\label{MasterEq}\\
&=& C_S(Q^2/\mu^2,{\bar g}(\mu^2),\alpha)\langle \gamma(p)|
R_S(\mu^2)|\gamma(p)\rangle \nonumber\\
&+& 
 C_{NS}(Q^2/\mu^2,{\bar g}(\mu^2),\alpha)\langle \gamma(p)|
R_{NS}(\mu^2)|\gamma(p)\rangle~. \nonumber
\end{eqnarray}
Here $C_S$ and $C_{NS}$ are the coefficient functions 
corresponding to the currents $R_{S}^{\sigma}$ and $R_{NS}^{\sigma}$,
respectively, and $\langle \gamma(p)|R_i(\mu^2)|\gamma(p)\rangle$ 
with $i=S, NS$ are the photon matrix elements of these quark axial currents.
We choose the renormalization point $\mu$ at
$\mu^2=P^2$.

The coefficient functions are given by
\begin{eqnarray}
&&\hspace{-0.8cm}C_i(Q^2/P^2,{\bar g}(P^2),\alpha) \nonumber\\
&=&\!\!\exp\left[\int_{{\bar g}(Q^2)}^{{\bar
g}(P^2)}dg'\frac{{\gamma}_i(g')}{\beta(g')}\right]C_i(1,{\bar g}(Q^2),\alpha),
\end{eqnarray}
where $i=S, NS$ and ${\gamma}_i(g)$ is the anomalous dimension of the quark axial current
$R_i^\sigma$ and $\beta(g)$ is the QCD $\beta$-function. 
We expand ${\gamma}_i(g)$ in powers of $g$ as
\begin{eqnarray}
&&\hspace{-0.8cm}\gamma_i(g)=\gamma_i^{(0)}\frac{g^2}{16\pi^2}+\gamma_i^{(1)}(\frac{g^2}{16\pi^2})^
2\nonumber \\
&&\qquad \qquad +\gamma_i^{(2)}(\frac{g^2}{16\pi^2})^3+{\cal O}(g^8)~. \label{AnomalousDimension}
\end{eqnarray}
Since the nonsinglet quark axial current $R_{NS}^\sigma$ is conserved in the massless
limit,  it undergoes no renormalization, and thus  we have 
\begin{equation}
\gamma_{NS}^{(0)}=\gamma_{NS}^{(1)}=\gamma_{NS}^{(2)}=\cdots =0~.
\end{equation}
On the other hand, the singlet axial current $R_{S}^\sigma$ has 
a non-zero anomalous dimension $\gamma_S(g)$ due to the axial anomaly. 
At the one-loop level, we know $\gamma_{S}^{(0)}=0$.  
The two-loop~\cite{JK} and three-loop~\cite{Larin1} 
results are:
\begin{eqnarray}
\gamma_{S}^{(1)}\!\! &=&\!\! 12C_F n_f,\\
\gamma_{S}^{(2)}\!\! &=&\!\! \Bigl(\frac{284}{3}C_FC_A-36C_F^2\Bigr)n_f-\frac{8}{3}C_Fn_f^2,
\end{eqnarray}
with $C_F=\frac{4}{3}$ and $C_A=3$. 
The result $\gamma_{S}^{(2)}$ was obtained in the $\overline{\rm MS}$ scheme 
with a definition of the $\gamma_5$-matrix as
\begin{equation}
\gamma_\mu\gamma_5=\frac{i}{3!}\epsilon_{\mu\rho\sigma\tau}\gamma^\rho\gamma^\sigma\gamma^\tau, 
\label{gamma5}
\end{equation} 
being used.

The $\beta$ function has been calculated up to the four-loop level 
in the $\overline{\rm MS}$ scheme~\cite{vRVL,Czakon}. 
For numerical analysis, we will use later the QCD running coupling constant $\alpha_s(Q^2)$ 
where the results up to the three-loop level are taken care of \cite{AlphaStrong}.  
But for the present  we only need the
$\beta$ function  up to the two-loop level:
\begin{eqnarray}
&&\hspace{-0.8cm}\mu \frac{\partial g}{\partial \mu}=\beta(g) \nonumber\\
&&=-\beta_0 \frac{g^3}{16\pi^2}-\beta_1
\frac{g^5}{(16\pi^2)^2}+\cdots~, \label{betaFunc}
\end{eqnarray}
with the $SU_C(3)$ value
\begin{eqnarray}
\beta_0&=&11-\frac{2}{3} n_f~, \nonumber\\
\beta_1&=&102-\frac{38}{3}n_f~.  
\end{eqnarray}
Using Eqs.(\ref{AnomalousDimension}) and (\ref{betaFunc}) we obtain
\begin{eqnarray}
&&\hspace{-0.8cm}\exp\left[\int_{{\bar
g}(Q^2)}^{{\bar g}(P^2)}dg'\frac{{\gamma}_S(g')}{\beta(g')}\right]\nonumber \\
&&\hspace{-0.8cm}= 1 + 
\frac{\gamma_S^{(1)}}{8\beta_0} 
 \Bigl(\frac{\alpha_s(Q^2)}{\pi}-\frac{\alpha_s(P^2)}{\pi}\Bigr)\\
&&\hspace{-0.6cm}+ \frac{1}{64\beta_0}\Bigl(\gamma_S^{(2)} -\gamma_S^{(1)}\frac{\beta_1}{\beta_0}  
\Bigr)\nonumber \\
&&\hspace{1.5cm} \times\Bigl[\Bigl( \frac{\alpha_s(Q^2)}{\pi}  \Bigr)^2- \Bigl(
\frac{\alpha_s(P^2)}{\pi} 
\Bigr)^2 
\Bigr]\nonumber\\
&&\hspace{-0.6cm}+ \frac{1}{128}\Bigl(\frac{\gamma_S^{(1)}}{\beta_0}   \Bigr)^2
\biggl( \frac{\alpha_s(Q^2)}{\pi} - \frac{\alpha_s(P^2)}{\pi}  \biggr)^2
+{\cal O}(\alpha_s^3). \nonumber    
\end{eqnarray}

We already have results of the singlet coefficient function $C_{S}(1,{\bar g}(Q^2),\alpha)$ 
calculated  up to the two-loop level~\cite{ZvN,Larin2} 
and  the nonsinglet coefficient function $C_{NS}(1,{\bar g}(Q^2),\alpha)$ 
up to the three-loop level~\cite{LV1}. Both  were calculated  
in the $\overline{\rm MS}$ scheme and with the definition of the $\gamma_5$-matrix 
in (\ref{gamma5}).
Taking up to the two-loop level, we have
\begin{eqnarray}
&&\hspace{-0.8cm}C_S(1,{\bar
g}(Q^2),\alpha)/\langle e^2\rangle=1-\frac{3}{4}C_F\frac{\alpha_s(Q^2)}{\pi}\nonumber\\
&& +
C_F\Bigl[\frac{21}{32} C_F -\frac{23}{16}C_A+(\frac{1}{2}\zeta_3+ \frac{13}{48}) n_f
\Bigr]\nonumber\\
&& \qquad \times\Bigl(\frac{\alpha_s(Q^2)}{\pi}\Bigr) ^2, \\
&&\hspace{-0.8cm}C_{NS}(1,{\bar g}(Q^2),\alpha)=
1-\frac{3}{4}C_F\frac{\alpha_s(Q^2)}{\pi}  \\
&&+C_F\Bigl( \frac{21}{32}C_F-
\frac{23}{16}C_A +\frac{1}{4}n_f\Bigr)\Bigl( \frac{\alpha_s(Q^2)}{\pi} \Bigr)^2.\nonumber
\end{eqnarray}
where $\zeta_3$ is the Riemann zeta-function ($\zeta_3=1.202056903\cdots$).

For $-p^2=P^2 \gg \Lambda^2$,  the photon matrix elements of the quark axial currents
can be calculated perturbatively and they are expressed in the form as  
\begin{equation}
\langle\gamma(p)|R_i(\mu^2=P^2)|\gamma(p)\rangle=
\frac{\alpha}{4\pi}A_i~,
\end{equation}
with $i=S, NS$, and 
\begin{equation}
A_i=A_i^{(0)}+\frac{{\overline g}^2(P^2)}{16\pi^2}A_i^{(1)}+
\Bigl(\frac{{\overline g}^4(P^2)}{16\pi^2}\Bigr)^2 A_i^{(2)}+\cdots.
\end{equation}
The leading terms $A_i^{(0)}$ are connected with Adler-Bell-Jackiw anomaly and are given
by\cite{MvN}
\begin{eqnarray}
A_S^{(0)}&=&-12n_f \langle e^2  \rangle~, \\
A_{NS}^{(0)}&=& -12n_f(\langle e^4  \rangle-\langle e^2  \rangle^2)~.
\end{eqnarray}
In our talk given at RADCOR05, we reported that 
for the next-leading and next-next-leading terms $A_i^{(1)}$ and 
$A_i^{(2)}$,  we have 
$A_S^{(1)}=A_{NS}^{(1)}=A_S^{(2)}=A_{NS}^{(2)}=0$
due to the nonrenormalization theorem~\cite{AB} for  the triangle anomaly. 
It is true that we obtain
\begin{equation}
A_S^{(1)}=A_{NS}^{(1)}=A_{NS}^{(2)}=0~,
\end{equation}
but we found after the workshop that $A_S^{(2)}$ has a non-vanishing value. 
 For the calculation of $A_S^{(2)}$, we need to 
evaluate the three-loop Feynman graphs. Instead we resort to the Adler-Bardeen theorem 
for the axial current $J_5^\mu(= R_{S}^{\mu}$ in (\ref{axialcurrent})),
\begin{equation}
\partial_\mu J_5^\mu=\frac{g^2}{16\pi^2}\frac{n_f}{2}G_{\mu\nu}{\tilde G}^{\mu\nu},\label{AdlerBardeen}
\end{equation} 
which holds in all orders in $\alpha_s$, and we evaluate, in the $\overline{\rm MS}$ scheme, 
the matrix element
\begin{equation}
{\epsilon_\lambda}^{\rho\sigma\tau}\langle 0|T[A_\rho^a\partial_\sigma A_\tau^a A_\mu(-p)
A_\nu(p)]|0 \rangle_{\rm amputated}~, \label{PMofGluon}
\end{equation}
in the two-loop level, where $A_\rho^a$ and $A_\mu$ are gluon and photon fields, 
respectively.  Details will be reported
elsewhere~\cite{SUU}.   We found 
\begin{equation}
A_S^{(2)}=24n_f\langle e^2  \rangle C_F \frac{n_f}{2}\Bigl(\frac{53}{3} -8 \zeta_3 \Bigr)~.
\end{equation}

Putting all these equations into Eq.(\ref{MasterEq}) with $\mu^2=P^2$, we finally
obtain the NNLO ($\alpha\alpha_s^2$) corrections to 
the first moment of $g_1^\gamma(x,Q^2,P^2)$:
\begin{eqnarray}
&&\hspace{-0.8cm}\int_0^1 dx g_1^\gamma(x,Q^2,P^2)/\Bigl( - \frac{3\alpha}{\pi} \Bigr)\nonumber\\
&&\hspace{-0.8cm} =\sum_i^{n_f} e_i^4
\left[1-\frac{\alpha_s(Q^2)}{\pi}\right]\nonumber\\
&&\hspace{-0.6cm} - \frac{2}{\beta_0} \Bigl(\sum_i^{n_f} e_i^2 \Bigr)^2
\left[\frac{\alpha_s(P^2)}{\pi}-\frac{\alpha_s(Q^2)}{\pi}\right] \nonumber\\
&&\hspace{-0.6cm} + \frac{2}{\beta_0} \Bigl(\sum_i^{n_f} e_i^2\Bigr)^2
\frac{\alpha_s(Q^2)}{\pi}
\left[\frac{\alpha_s(P^2)}{\pi}-\frac{\alpha_s(Q^2)}{\pi}\right] \nonumber\\
&&\hspace{-0.6cm} + \frac{1}{4\beta_0}
\Bigl(\frac{\beta_1}{\beta_0}-\frac{59}{3}+\frac{2}{9}n_f\Bigr) \Bigl(\sum_i^{n_f}
e_i^2\Bigr)^2\nonumber\\
&&\hspace{1.5cm} \times\left[\frac{\alpha_s^2(P^2)}{\pi^2}-\frac{\alpha_s^2(Q^2)}{\pi^2}\right]
\nonumber\\
 &&\hspace{-0.6cm} + \frac{2n_f}{\beta_0^2} \Bigl(\sum_i^{n_f} e_i^2\Bigr)^2
\left[\frac{\alpha_s(P^2)}{\pi}-\frac{\alpha_s(Q^2)}{\pi}\right]^2 \nonumber\\
&&\hspace{-0.6cm} - \Bigl(\frac{55}{12}-\frac{1}{3}n_f\Bigr) \sum_i^{n_f} e_i^4
\frac{\alpha_s^2(Q^2)}{\pi^2}\nonumber\\
&&\hspace{-0.6cm} + \Bigl(\frac{2}{3}\zeta_3+\frac{1}{36}\Bigr) \Bigl(\sum_i^{n_f} e_i^2\Bigr)^2
\frac{\alpha_s^2(Q^2)}{\pi^2}\nonumber\\
&&\hspace{-0.6cm} -\frac{1}{12} \Bigl(\frac{53}{3}-8\zeta_3\Bigr) \Bigl(\sum_i^{n_f}
e_i^2\Bigr)^2
\frac{\alpha_s^2(P^2)}{\pi^2}~, 
\label{finalResult}
\end{eqnarray}
where the third to 8th terms  are the NNLO contributions.
In the case of $n_f=4$, for an example, we have
\begin{eqnarray}
&&\hspace{-0.8cm}\int_0^1 dx g_1^\gamma(x,Q^2,P^2)\nonumber\\
&&\hspace{-0.8cm} =-
\frac{3\alpha}{\pi}\Biggl\{0.4198-0.1235\frac{\alpha_s(Q^2)}{\pi}
-0.2963\frac{\alpha_s(P^2)}{\pi}\nonumber\\
&&\hspace{-0.5cm} -0.02731\Bigl( \frac{\alpha_s(Q^2)}{\pi}  \Bigr)^2 +0.01185
\frac{\alpha_s(Q^2)}{\pi}\frac{\alpha_s(P^2)}{\pi}\nonumber\\
&&\hspace{1.5cm} -1.153\Bigl(
\frac{\alpha_s(P^2)}{\pi}  \Bigr)^2  \Biggr\}~.
\end{eqnarray}

To estimate the sizes of the NLO ($\alpha\alpha_s$) and NNLO ($\alpha\alpha_s^2$)
corrections compared to the LO ($\alpha$) term, we take, for instance, $Q^2=30$ and 100 ${\rm
GeV}^2$,  and $P^2=3~{\rm GeV}^2$. The corresponding values of $\alpha_s$ are 
obtained from Ref.\cite{AlphaStrong}. We get $\alpha_s(Q^2\!=\!30{\rm
GeV}^2)\!=\!0.2048$, $\alpha_s(Q^2\!=\!100{\rm
GeV}^2)\!=\!0.1762$, and $\alpha_s(P^2\!=\!3{\rm
GeV}^2)\!=\!0.3211$. The results are given in Table \ref{NNLOvsSUM}.

\begin{table*}[tb] 
\begin{center}
  \caption{The NLO and NNLO contributions relative to LO
           in the first moment of $g_1^\gamma(x,Q^2,P^2)$.}
  \begin{tabular}{c l ccc r@{.}l r@{.}l r@{.}l c}
    \hline
    \quad & & $Q^2/\mbox{GeV}^2$ & $P^2/\mbox{GeV}^2$ & LO
          & \multicolumn{2}{c}{NLO} & \multicolumn{2}{c}{NNLO}
          & \multicolumn{2}{c}{NNLO/(LO+NLO)}
          & \quad \\
    \hline
    & $n_f=3$ & \phantom{0}30 & 3 & 1 & -0&0816  & -0&0250 & -0&0272\\
    &         &           100 & 3 & 1 & -0&0766  & -0&0234 & -0&0254\\
    & $n_f=4$ & \phantom{0}30 & 3 & 1 & -0&0913  & -0&0288 & -0&0317\\
    &         &           100 & 3 & 1 & -0&0886  & -0&0287 & -0&0315\\
    & $n_f=5$ & \phantom{0}30 & 3 & 1 & -0&0986  & -0&0309 & -0&0343\\
    &         &           100 & 3 & 1 & -0&0977  & -0&0319 & -0&0354\\
    \hline
  \end{tabular}
\label{NNLOvsSUM}
\end{center}
\end{table*}

\section{Summary}

We have investigated  the next-to-next-to-leading order ($\alpha \alpha_s^2$) corrections to the 
first moment of the  polarized virtual photon 
structure function $g_1^\gamma(x,Q^2,P^2)$ in the kinematical region
$Q^2\gg P^2 \gg \Lambda^2$ in QCD. All the necessary information on 
the coefficient functions and anomalous dimensions corresponding to 
the quark axial currents has been already known, except for 
the three-loop-level photon matrix element (the finite term) of the flavour singlet quark axial 
current $R^\mu_S$. Instead of evaluating the  relevant  three-loop Feynman diagrams, we resort to  
the Adler-Bardeen theorem  for the axial current (\ref{AdlerBardeen}). Then 
calculation reduces to the one in the two-loop level. We evaluate in effect the two-loop 
diagrams for the photon matrix element of the gluon operator (\ref{PMofGluon}).

The  $\alpha \alpha_s^2$ corrections are found to be about 3\%
of the sum of the leading ($\alpha$) and the next-to-leading ($\alpha\alpha_s$) contributions,
when $Q^2=30\sim 100 {\rm GeV}^2$ and $P^2=3 {\rm GeV}^2$.



\begin{thebibliography}{9}
\bibitem{Kraw}M. Krawczyk, Talk given at Photon 2000,
         AIP Conf. Proc. {\bf 571}~(2001)~3.

\bibitem{Nisi}R. Nisius, Talk at Photon 2001,
         Proc. Photon 2001.

\bibitem{Klas}M. Klasen, 
         {\sl Rev.~Mod.~Phys.}~{\bf 74}~(2002)~1221. 

\bibitem{Schi}I. Schienbein, 
         {\sl Ann. Phys.}~{\bf 301}~(2002)~128.

\bibitem{ET}
        A.~V.~Efremov and O.~V.~Teryaev, JINR Report NO. E2-88-287, Dubna, 
        1988; {\sl Phys.~Lett.} {\bf B240}, 200 (1990).
\bibitem{BASS}
          S.~D.~Bass, {\sl Int.~J.~Mod.~Phys.} {\bf A7}, 6039 (1992).
\bibitem{NSV}
          S.~Narison, G.~M.~Shore and G.~Veneziano,
           {\sl Nucl.~Phys.} {\bf B391}, 69 (1993); 
          G.~M.~Shore and G.~Veneziano, {\sl Mod.~Phys.~Lett.}
           {\bf A8}, 373 (1993);
          G.~M.~Shore and G.~Veneziano, {\sl Nucl.~Phys.} {\bf B381}, 23 (1992);\\
          G.~M.~Shore; {\sl Nucl.~Phys.} {\bf B712}, 411 (2005).


\bibitem{FS}
          A.~Freund and L.~M.~Sehgal, {\sl Phys.~Lett.} {\bf B341}, 90 (1994).
\bibitem{BBS}
          S.~D.~Bass, S.~J.~Brodsky and I.~Schmidt,
           {\sl Phys.~Lett.} {\bf B437}, 424 (1998).

\bibitem{KS}
         K.~Sasaki, {\sl Phys. Rev.} {\bf D22} (1980) 2143; 
          {\sl Prog. Theor. Phys. Suppl. }~{\bf 77} (1983) 197.


\bibitem{SV}
         M. Stratmann and W. Vogelsang, {\it Phys. Lett.} {\bf B386}, 
         (1996) 370 .

\bibitem{GRS}
          M.~Gl{\"u}ck, E.~Reya and C.~Sieg, {\sl Phys.~Lett.}
         {\bf B503}, 285 (2001);\\  {\sl Eur. Phys. J.} {\bf C20}, 271 (2001).
\bibitem{SU1}
       K.~Sasaki and T.~Uematsu, {\sl Phys. Rev.} {\bf D59} (1999)114011.
\bibitem{SU2}
       K.~Sasaki and T.~Uematsu, 
       {\sl Phys. Lett.} {\bf B473} (2000) 309; {\sl Eur.~Phys.~J.} {\bf C20} 
       (2001) 283.


\bibitem{UW}
     T.~Uematsu and T.~F.~Walsh, {\sl Phys. Lett.}{\bf 101B} (1981) 263, 
{\sl Nucl.~Phys.} {\bf B199} (1982) 93.

\bibitem{Collins}
       J. C. Collins, Renormalization (Cambridge University Press, Cambridge, 1987)
 p.321.

\bibitem{AB}
     S.~L.~Adler and W.~A.~Bardeen, {\sl Phys.~Rev.} {\bf 182} (1969) 1517;~ 
  W.~A.~Bardeen, {\sl Phys.~Rev.} {\bf 184} (1969) 1848.

\bibitem{JK}
     J.~Kodaira, {\sl Nucl.~Phys.} {\bf B165} (1980) 129.


\bibitem{Larin1}
     S.A.~Larin, {\sl Phys.~Lett.} {\bf B303} (1993) 113.


\bibitem{vRVL}
     T.~van Ritbergen, J.A.M.~Vermaseren and S.A.~Larin, {\sl Phys.~Lett.} {\bf B400}
(1997) 379.

\bibitem{Czakon}
     M.~Czakon, {\sl Nucl.~Phys.} {\bf B710} (2005) 485.

\bibitem{AlphaStrong}
   I.~Hinchliffe: http://www-thory.lbl.gov/ \\ \~{} ianh/alphat/alpha.htm


\bibitem{ZvN}
     E.~B.~Zijlstra and W.~L.~van~Neerven, {\sl Nucl.~Phys.} {\bf B417} 
     (1994) 61.

\bibitem{Larin2}
     S.A.~Larin, {\sl Phys.~Lett.} {\bf B334} (1994) 192.

\bibitem{LV1}
     S.A.~Larin and J.A.M.~Vermaseren {\sl Phys.~Lett.} {\bf B259} (1991) 345.

\bibitem{MvN}
     R.~Mertig and W.~L.~van Neerven, {\sl Z.~Phys.} {\bf C70} (1996) 637.

\bibitem{SUU}
     K.~Sasaki, T.~Ueda and T.~Uematsu, in preparation.





\end{thebibliography}
\end{document}